\documentclass{article}
\usepackage{amsmath, amssymb, amsfonts}
\title{Anisotropic fluid with time dependent viscosity coefficients}
\author{Hristu Culetu, \\Ovidius University, Dept.of Physics, \\B-dul Mamaia 124, 8700 Constanta, Romania, \\e-mail : hculetu@yahoo.com}

\begin{document}
\numberwithin{equation}{section}
\pagenumbering{arabic}
\maketitle
\newcommand{\fv}{\boldsymbol{f}}
\newcommand{\tv}{\boldsymbol{t}}
\newcommand{\gv}{\boldsymbol{g}}
\newcommand{\OV}{\boldsymbol{O}}
\newcommand{\wv}{\boldsymbol{w}}
\newcommand{\WV}{\boldsymbol{W}}
\newcommand{\NV}{\boldsymbol{N}}
\newcommand{\hv}{\boldsymbol{h}}
\newcommand{\yv}{\boldsymbol{y}}
\newcommand{\RE}{\textrm{Re}}
\newcommand{\IM}{\textrm{Im}}
\newcommand{\rot}{\textrm{rot}}
\newcommand{\dv}{\boldsymbol{d}}
\newcommand{\grad}{\textrm{grad}}
\newcommand{\Tr}{\textrm{Tr}}
\newcommand{\ua}{\uparrow}
\newcommand{\da}{\downarrow}
\newcommand{\ct}{\textrm{const}}
\newcommand{\xv}{\boldsymbol{x}}
\newcommand{\mv}{\boldsymbol{m}}
\newcommand{\rv}{\boldsymbol{r}}
\newcommand{\kv}{\boldsymbol{k}}
\newcommand{\VE}{\boldsymbol{V}}
\newcommand{\sv}{\boldsymbol{s}}
\newcommand{\RV}{\boldsymbol{R}}
\newcommand{\pv}{\boldsymbol{p}}
\newcommand{\PV}{\boldsymbol{P}}
\newcommand{\EV}{\boldsymbol{E}}
\newcommand{\DV}{\boldsymbol{D}}
\newcommand{\BV}{\boldsymbol{B}}
\newcommand{\HV}{\boldsymbol{H}}
\newcommand{\MV}{\boldsymbol{M}}
\newcommand{\be}{\begin{equation}}
\newcommand{\ee}{\end{equation}}
\newcommand{\ba}{\begin{eqnarray}}
\newcommand{\ea}{\end{eqnarray}}
\newcommand{\bq}{\begin{eqnarray*}}
\newcommand{\eq}{\end{eqnarray*}}
\newcommand{\pa}{\partial}
\newcommand{\f}{\frac}
\newcommand{\FV}{\boldsymbol{F}}
\newcommand{\ve}{\boldsymbol{v}}
\newcommand{\AV}{\boldsymbol{A}}
\newcommand{\jv}{\boldsymbol{j}}
\newcommand{\LV}{\boldsymbol{L}}
\newcommand{\SV}{\boldsymbol{S}}
\newcommand{\av}{\boldsymbol{a}}
\newcommand{\qv}{\boldsymbol{q}}
\newcommand{\QV}{\boldsymbol{Q}}
\newcommand{\ev}{\boldsymbol{e}}
\newcommand{\uv}{\boldsymbol{u}}
\newcommand{\KV}{\boldsymbol{K}}
\newcommand{\ro}{\boldsymbol{\rho}}
\newcommand{\si}{\boldsymbol{\sigma}}
\newcommand{\thv}{\boldsymbol{\theta}}
\newcommand{\bv}{\boldsymbol{b}}
\newcommand{\JV}{\boldsymbol{J}}
\newcommand{\nv}{\boldsymbol{n}}
\newcommand{\lv}{\boldsymbol{l}}
\newcommand{\om}{\boldsymbol{\omega}}
\newcommand{\Om}{\boldsymbol{\Omega}}
\newcommand{\Piv}{\boldsymbol{\Pi}}
\newcommand{\UV}{\boldsymbol{U}}
\newcommand{\iv}{\boldsymbol{i}}
\newcommand{\nuv}{\boldsymbol{\nu}}
\newcommand{\muv}{\boldsymbol{\mu}}
\newcommand{\lm}{\boldsymbol{\lambda}}
\newcommand{\Lm}{\boldsymbol{\Lambda}}
\newcommand{\opsi}{\overline{\psi}}
\renewcommand{\tan}{\textrm{tg}}
\renewcommand{\cot}{\textrm{ctg}}
\renewcommand{\sinh}{\textrm{sh}}
\renewcommand{\cosh}{\textrm{ch}}
\renewcommand{\tanh}{\textrm{th}}
\renewcommand{\coth}{\textrm{cth}}

\begin{abstract}
A spacetime endowed with an anisotropic fluid is proposed for the interior of the event horizon of a black hole. The geometry has an instantaneous Minkowski form and is a solution of Einstein's equations with a stress tensor on the r.h.s. obeying all the energy conditions.

The interior fluid is compressible, with time dependent shear and bulk viscosity coefficients. The energy density $\rho$ and the ''radial'' pressure $p$ obey the equation of state  $p + \rho = 0$ (as for dark energy), with no pressures on $\theta-$ and $\phi-$ directions. However, the angular components of the viscous part of the stress tensor are nonvanishing and equals the energy density of the fluid.
\end{abstract}


\section{Introduction}
 Many authors expressed the idea that the interior of a black hole can be thought of as an anisotropic collapsing cosmology \cite{HL}. The interior of a Schwarzschild black hole may be considered as a homogeneous anisotropic cosmology of the Kantowski - Sachs family \cite{KS} \cite{RB}. As Brehme has noticed, when $r < 2m$, a remarkable change occurs in the nature of spacetime : the temporal coordinate for the outside observer becomes a spatial coordinate for the inside observer, the interior geometry being time dependent. In addition, the area of the two-surface of constant ''radial'' and time coordinates increases at a rate proportional with time.
 
 The mass $m$ of the black hole appears only at the moment $t = 0$, distributed uniformly along the ''radial'' coordinate $z$ inside the horizon \cite{RB}. 
 
 In a different context, Cattoen et al. \cite{CFV} showed that the pressure anisotropy, implicit in the Mazur - Mottola model \cite{MM1}, is a necessity for all ''gravastars''-like (gravitational vacuum stars) objects (see also Dymnikova \cite{ID}) . A phase transition near the event horizon of a black hole (with an equation of state $p = -\rho < 0$) is produced. Moreover, Chapline \cite{GC} suggested that the spacetime undergoes a continuous phase transition where General Relativity  predicts there should be an event horizon (a quantum critical surface), called ''dark energy star''. The interior of the compact object has a much larger vacuum energy than the cosmological vacuum energy outside the horizon, with no singularity in the interior.
 
 We conjecture in the present article that the interior of a black hole is filled with an anisotropic fluid, its equation of state being given by $p(t) = -\rho(t)$, where the negative ''radial'' pressure $p$ and the energy density $\rho$ are time dependent.
 
 Following an idea of Doran et al. \cite{DLC} we chose the line element inside the horizon to be Minkowskian for constant angular variables (i.e., the mass of the black hole is proportional with time). The corresponding stress tensor obeys all the energy conditions : the weak, strong and the dominant one. We show that the fluid is characterized by time dependent coefficients of shear and bulk viscosities, a follow up of the pressure and energy density time dependance. 
 We found that the Raychaudhuri equation is fulfilled for a congruence of particles having timelike velocities $u^{\alpha}$ as they fall under their own gravity. We compute also the anisotropic stress tensor and show that the angular components of the viscous part of it are equal with the energy density $\rho$.

 \section {The time dependent interior spacetime}
 We know that the geometry inside the horizon of a Schwarzschild black hole is dynamic since the radial coordinate becomes timelike and the metric is time dependent. 
 It is given by 
 \begin{equation}
 ds^{2} = -\left(\frac{2m}{t}-1\right)^{-1} dt^{2} + \left(\frac{2m}{t}-1\right) dz^{2} + t^{2} d\Omega^{2} 
 \label {2.1}
 \end{equation}
 where $m$ is the mass of the black hole, $t \in (0, 2m)$ and $z$ plays the role of the radial coordinate, with $-\infty < z < +\infty$ and $d\Omega^{2} = d\theta^{2} + sin^{2} \theta d\phi^{2}$. 
  Throughout the paper we take the velocity of light $c = 1$ and Newton's constant G = 1.

 We know that the parameter $m$ is measured at spatial infinity where the Newtonian approximation is valid. Therefore, its physical meaning may be different inside the event horizon. Our conjecture in this paper is that $m$ represents, from the point of view of an interior observer, the Yu - Caldwell \cite{YC} quasi local energy $(QLE)_{in} = r$, which matches their expression of the QLE for a black hole viewed by an exterior observer (see also \cite{LSY})
 \begin {equation}
 (QLE)_{out} = r \left(1 - \sqrt{1-\frac{2m}{r}}\right).
 \label {2.2}
 \end{equation}
  We have however remind that $r$ is timelike inside the horizon. That means $m = t$, i.e. $m$ is time dependent \cite{DLC} \cite{LSY} \cite{DV} whereas QLE is position dependent (the quasilocal energy of a black hole is well - known to be observer dependent). Let us note that a similar time dependence was obtained by Harada et al. \cite{HMC} for the mass function of the Kantowski - Sachs spacetime, in self-similar cordinates.\\Therefore, the metric (2.1) acquires now, for $m(t) = t$, an instantaneous Minkowski form \cite{DLC}
 
\begin{equation}
ds^{2} = -dt^{2} + dz^{2} + t^{2} d\Omega^{2} 
\label{2.3}
\end{equation}
The line element (2.3) has a curvature singularity at $t = 0$ since the scalar curvature $R_{\alpha}^{\alpha} = 4/t^{2}$ is infinite there. In addition, $t$ is no longer restricted to the previous interval, taking any real value.
However, a  geodesic particle with constant angular coordinates moves exactly as in flat space. 

 In order to be a solution of Einstein's equations, we must have an energy - momentum tensor ( a source ) on the r.h.s. In the spirit of Lobo \cite{FL} and Viaggiu \cite{SV} we consider the spacetime (2.3) to be endowed with an anisotropic fluid with the stress tensor 
 \begin{equation}
 T_{\mu}^{\nu} = \rho u_{\mu} u^{\nu} + p s_{\mu} s^{\nu}
 \label{2.4}
 \end{equation}
 where $u_{\mu} = (-1,0,0,0)$ is the fluid 4 - velocity (the components are in order t, z, $\theta$, $\phi$), $s_{\mu} = (0,1,0,0)$ is the unit spacelike vector in the direction of anisotropy (with $u_{\alpha} u^{\alpha} = -1$, $s_{\alpha}s^{\alpha} = 1~~and~~s_{\alpha} u^{\alpha} = 0$). Hence, we have only one principal pressure $p$ and the transverse pressures measured in the orthogonal direction to $s^{\mu}$ are vanishing. 
 
 With (2.4) on the r.h.s. of the Einstein equations, one obtains, for the energy density and pressure on the $z$ - direction
 \begin{equation}
 \rho(t) = -T_{t}^{t} = \frac{1}{4 \pi t^{2}},~~~p(t) = T_{z}^{z} = -\frac{1}{4 \pi t^{2}},~~T_{\theta}^{\theta} = T_{\phi}^{\phi} = 0.
 \label{2.5}
 \end{equation}
 (The convention $R_{\alpha \beta} = \partial_{\nu} \Gamma_{\alpha \beta}^{\nu}-...$ has been used). We note that $p = -\rho$, as for dark energy stars \cite{LSY} \cite{GC}.
 
 It is an easy task to check that the weak, strong and dominant energy conditions for $T_{\mu \nu}$ are obeyed. We choose  $\xi^{\alpha} = (\xi^{(t)},\xi^{(z)},0,0)$ as any future directed  timelike vector with only two components, for simplicity. Since
 \begin{equation}
 \xi^{\alpha} \xi_{\alpha} = -(\xi^{t})^{2} + (\xi^{z})^{2} < 0     
 \label{2.6}
 \end{equation}
 we obtain
 \begin{equation}
 T_{\mu \nu} \xi^{\mu} \xi^{\nu} = \rho \left[(\xi^{t})^{2}- (\xi^{z})^{2}\right] > 0.
 \label{2.7}
 \end{equation}
 Hence, the weak energy condition is satisfied.
 For the  strong energy condition , we have
 \begin{equation}
 (T_{\mu \nu} - \frac{1}{2}g_{\mu \nu}T) \xi^{\mu} \xi^{\nu} = R_{\mu \nu} \xi^{\mu} \xi^{\nu} = 0
 \label{2.8}
 \end{equation}
 because in our spacetime $R_{tt} = R_{zz} = 0$
The energy - momentum 4 - current density of the fluid \cite{RW}
 \begin{equation}
 - T_{\alpha}^{\beta} \xi^{\alpha} = \rho \xi^{\beta} 
 \label{2.9}
 \end{equation}
 is a timelike vector ($\rho$ is positive) and the dominant energy condition is also satisfied.

 \section{The anisotropic stress tensor}
  Let us consider the general expression of the stress-energy tensor for a compressible, viscous fluid with heat conduction \cite{KM} 
  \begin{equation}
  T_{\alpha \beta} = \rho u_{\alpha} u_{\beta}+p h_{\alpha \beta} + \pi_{\alpha \beta} +q_{\alpha}u_{\beta}+ q_{\beta}u_{\alpha}
  \label{3.1}
 \end{equation}
 where the viscous part of the stress tensor is given by
 \begin{equation}
 \pi_{\alpha \beta} = 2 \eta \sigma_{\alpha \beta} + \zeta \Theta h_{\alpha \beta}.
 \label{3.2}
 \end{equation}
  $\Theta$ is the expansion of the fluid worldlines (the rate of increasing of a fluid volume element) given by the divergence of $u_{\alpha}$
\begin{equation}
\Theta = \nabla _{\alpha}u^{\alpha} ,
\label{3.3}
\end{equation}
$h_{\alpha \beta} = g_{\alpha \beta}+u_{\alpha}u_{\beta}$ is the projection tensor onto the direction perpendicular to $u_{\alpha}$ and $\sigma_{\alpha \beta}$ expresses the distorsion of the fluid in shape without change in volume \cite{ND}. It is orthogonal to $u^{\alpha} (i.e., u^{\alpha} \sigma_{\alpha \beta}= 0$), tracefree and may be computed from
\begin{equation}
\sigma_{\alpha \beta} = \frac{1}{2}(h_{\beta}^{\mu} \nabla_{\mu} u_{\alpha}+ h_{\alpha}^{\mu} \nabla_{\mu} u_{\beta})-\frac{1}{3} \Theta h_{\alpha \beta}+ \frac{1}{2}(a_{\alpha} u_{\beta} + a_{\beta} u_{\alpha}).
\label{3.4}
\end{equation}
$\eta ~and ~\zeta$ are the coefficients of shear and, respectively, bulk viscosity. $q_{\alpha}$ is the heat flux 4-vector
\begin{equation}
q^{\alpha} = - \kappa T_{\nu \beta}u^{\nu} h^{\alpha \beta}
\label{3.5}
\end{equation}
with $\kappa$ - the thermal conductivity and $a_{\alpha}$ is the acceleration due to nongravitational forces.
\begin{equation}
a_{\alpha} = u^{\beta} \nabla_{\beta} u_{\alpha}.
\label{3.6}
\end{equation}
Before to compare the tensors (2.4) and (3.1), we note that (2.4) is a particular case of the locally anisotropic fluid with nonzero transversal pressure $p_{t}$ and a stress tensor given by \cite{SV} \cite{AB}
\begin{equation}
\tau_{\alpha \beta} = (\rho + p_{t})u_{\alpha} u_{\beta} + p_{t} g_{\alpha \beta}+ (p-p_{t})s_{\alpha} s_{\beta}
\label{3.7}
\end{equation}
 We see that a vanishing tangential pressure in (3.7) leads to (2.4), with $p$ as the pressure along the anisotropy.
 
 The geometry (2.3) has the following nonzero Christoffel symbols :
 \begin{equation}
 \Gamma_{\theta \theta}^{t} = t,~~\Gamma_{t \theta}^{\theta}= \Gamma_{t \phi}^{\phi}=\frac{1}{t},~~\Gamma_{\theta \phi}^{\phi} = cot \theta,~~\Gamma_{\phi \phi}^{t} = t sin^{2}\theta,~~\Gamma_{\phi \phi}^{\theta} = -sin\theta cos\theta.
 \label{3.8}
 \end{equation}
 Therefore, the nonzero components of the Ricci tensor are given by  \cite{HC1}
 \begin{equation}
 R_{\theta}^{\theta} = R_{\phi} ^{\phi}= \frac{2}{t^{2}}
 \label{3.9}
 \end{equation}
 By means of (3.8) one obtains, from (3.3)
 \begin{equation}
 \Theta = \frac{2}{t},~~~\dot{\Theta} \equiv \frac{d\Theta}{d \tau}= -\frac{2}{t^{2}}
 \label{3.10}
 \end{equation}
 We observe that the expansion scalar is not invariant at a time reversal. It is negative for $t < 0$ (ingoing geodesics) and positive for $t > 0$ (outgoing geodesics), being singular at $t = 0$, where the curvature is infinite.
 
 The nonzero components of the shear tensor are 
 \begin{equation}
 \sigma_{z}^{z} = \frac{-2}{3t},~~~\sigma_{\theta}^{\theta}=\sigma_{\phi}^{\phi}= \frac{1}{3t}
 \label{3.11}
 \end{equation}
 with $h^{\alpha \beta} \sigma_{\alpha \beta} = 0$. From (3.5) and (3.6) we find that $q_{\alpha} =0~~and ~~a_{\alpha} = 0$ (we have a congruence of timelike geodesics). The same conclusion is valid for the vorticity (rotation without change in shape)
\begin{equation}
\omega_{\alpha \beta} \equiv \frac{1}{2}(h_{\beta}^{\mu} \nabla_{\mu}u_{\alpha}- h_{\alpha}^{\mu} \nabla_{\mu} u_{\beta}) + \frac{1}{2} (a_{\alpha}u_{\beta} - a_{\beta} u_{\alpha}) = 0.
\label{3.12}
\end{equation}
Using (3.9) and (3.10), it is an easy task to check that the Raychaudhuri equation \cite{ND} \cite{ANK}
\begin{equation}
\dot{\Theta} - \nabla_{\alpha} a^{\alpha}+ 2(\sigma^{2}- \omega^{2})+ \frac{1}{3} \Theta^{2} = R_{\alpha \beta} u^{\alpha} u^{\beta}
\label{3.13}
\end{equation}
is obeyed. We have here $2\sigma^{2} \equiv \sigma_{\alpha \beta} \sigma^{\alpha \beta}= 2/3t^{2}~~and~~2 \omega^{2} \equiv \omega_{\alpha \beta} \omega^{\alpha \beta} =0$.

We are now in a position to see whether the stress tensor (2.4) may be put in the general form (3.1), with the identification of the parameters $\eta~~and ~~\varsigma$. By a direct inspection, we conclude that we should have :
\begin{equation}
2\eta(t) = 3\zeta(t) = \frac{1}{4 \pi t}.
\label{3.14}
\end{equation}
(Using the fundamental constants, we obtain in fact $c^{4}/4 \pi G t$).
The fact that $2\eta = 3\zeta$ ( Stokes' relation) is known from the kinetic theory of gases \cite{FP}.

The time dependence of the coefficients of viscosity might be justified on the basis of Landau's and Lifshitz's remark \cite{LL} that in general the coefficients of viscosity may depend on the pressure of the fluid which, in our situation is a function of time. A similar time dependence have been obtained by Brevik et al. \cite{BNOV} in their study on the anisotropic Kasner metric which is valid for the very early Universe when viscosity comes into play whenever there are fluid sheets sliding with respect to each other.

We are now in a position to write down the components of the viscous contribution $\pi_{\alpha \beta}$ to the stress tensor $T_{\alpha \beta}$ 
\begin{equation}
\pi_{t}^{t} = \pi_{z}^{z} = 0,~~~~\pi_{\theta}^{\theta} = \pi_{\phi}^{\phi} = \frac{1}{4\pi t^{2}}
\label{3.15}
\end{equation}

The angular components of $\pi_{\alpha}^{\beta}$  and $\rho$ and $p$ from (2.5) are singular at $t = 0$, when the curvature of spacetime (2.3) is also infinite. Their time dependence is similar with that obtained by Maziashvili \cite{MM2} for the background energy density of the Universe (dark energy) (see also \cite{HC2})
\begin{equation}
\rho \cong \frac{1}{t_{P}^{2}t^{2}},
\label{3.18}
\end{equation}
where $t_{P}$ is the Planck time. He noticed that the existence of time $t$ fluctuating with the amplitude $\delta t$ implies the energy $(\delta t)^{-1}$ is distributed uniformly over the volume $t^{3}$. We stress that the Planck constant does not appear in the expression (3.16) (keeping track of all fundamental constants, we have in fact $\rho \cong c^{2}/Gt^{2}$). 

\section{Conclusions}
 A black hole interior filled with a compressible, anisotropic fluid is conjectured in this paper. The coefficients of shear and bulk viscosities are time dependent, a property which is justified by the fact that, in general $\eta$ and $\zeta$ may be functions of the pressure of the fluid which in our model is time dependent.
 
  The equation of state is $p + \rho = 0$ remind us of many dark energy models with a similar behaviour. In addition, the fact that $p$ and $\rho$ are proportional to $1/t^{2}$ resembles Maziashvili's model on the energy density of the Universe. Our geometry (2.3) has infinite scalar curvature on the hypersurface $t = 0$ and, therefore, the parameters $\rho, ~p, ~\eta, ~and ~\varsigma $ are also singular there.
 
 We have also checked that the Raychaudhuri equation for a congruence  of timelike geodesics is satisfied.


\begin{thebibliography} {25}
\bibitem {HL}
W.A Hiscock, S.L. Larson, Phys.Rev. D56, 3571 (1997).
\bibitem {KS}
R.Kantowski, R.K. Sachs, J.Math.Phys. (N.Y.) 7,443 (1966). (1966).
\bibitem {RB}
R.W. Brehme, Am.J.Phys.45,5, 423 (1977).
\bibitem {CFV}
C.Cattoen, T.Faber, M.Visser, gr-qc/0505137.
\bibitem{MM1}
P.O. Mazur and E.Mottola, gr-qc/0109035.
\bibitem {ID}
I. Dymnikova, Gen. Relat. Grav. 24, 235 (1992); Int.J.Mod.Phys. D12, 1015 (2003); I. Dymnikova and E. Galaktionov, Phys.Lett. B645, 358 (2007).
\bibitem {GC}
G.Chapline, astro-ph/0503200.
\bibitem {DLC}
R.Doran, F.S.N. Lobo, P.Crawford, gr-qc/0609042.
\bibitem {YC}
P.P. Yu and R.R. Caldwell, ArXiv 0801.3683 [gr-qc].
\bibitem{LSY}
A.P. Lundgren, B.S. Schmekel, J.W. York, Jr., gr-qc/0610088.
\bibitem{DV}
D.N. Vollick, hep-th/0102187.
\bibitem{HMC}
T. Harada et al., ArXiv 0707.0528 [gr-qc].
\bibitem{FL}
F.S.N. Lobo, gr-qc/0508115. 
\bibitem{SV}
S. Viaggiu, gr-qc/0603036.
\bibitem{RW}
R. Wald, General Relativity, The University of Chicago Press (1984).
\bibitem{KM}
T. Koivisto and D.F. Mota, ArXiv 0801.3676 [astro-ph].
\bibitem{ND}
N.Dadhich, gr-qc/0511123.
\bibitem{ANK}
A.Dasgupta, H.Nandan and S.Kar, 0709.0582 [ph-physics.class.].
\bibitem{BNOV}
I. Brevik et al., ArXiv hep-th/0401073.
\bibitem{AB}
A. DeBenedictis et al., gr-qc/0511097.
\bibitem{HC1}
H.Culetu, hep-th/0703168.
\bibitem{FP}
J.Florea, V.Panaitescu, Mecanica fluidelor, Editura Didactica si Pedagogica, Bucuresti (1979) (in Romanian).
\bibitem{LL}
L.Landau, E.M. Lifshitz, Mecanique de fluides, Editions Mir, Moscou (1971), tome 6. 
\bibitem{MM2}
M. Maziashvili, gr-qc/0612110.
\bibitem{HC2}
H. Culetu, hep-th/0701255.


\end{thebibliography}
\end{document}